# Anomalous tunneling characteristic of Weyl semimetals with tilted energy dispersion

Can Yesilyurt,[1,*] Zhuo Bin Siu,[1] Seng Ghee Tan,[1,2] Gengchiau Liang,[1] Shengyuan A. Yang,[3] and Mansoor B. A. Jalil[1,*]

[1]*Electrical and Computer Engineering, National University of Singapore, Singapore 117576, Republic of Singapore*

[2]*Data Storage Institute, Agency of Science, Technology and Research (A*Star), Singapore 138634, Republic of Singapore*

[3]*Research Laboratory for Quantum Materials, Singapore University of Technology and Design, Singapore 487372, Singapore*

Weyl semimetal is a recently discovered state of quantum matter, which generally possesses tilted energy dispersion. Here, we investigate the electron tunneling through a Weyl semimetal p-n-p junction. The angular dependence of electron tunneling exhibits an anomalous profile such that perfect transmission angles are shifted along the direction of the tilt. Coupling of the tilted dispersion and electrical potential within the barrier region gives rise to a transverse momentum shift, which is analogous to the transverse Lorentz displacement induced by magnetic barriers.

Electrons that encounter an electrical potential barrier undergo both refraction and reflection, a behavior analogous to ray optics[1,2]. Controlling electrons in condensed matter systems via these processes has paved the way for unique electron optics applications[1,3,4]. For Dirac-type electrons, the angular dependence of transmission probability, and the collimation as well as refraction of electrons has been investigated previously[5], particularly in the context of graphene. These studies showed that electrons obey the electronic analogue of Snell's law, where the refractive index is replaced by the electron's energy. Therefore, changing the carrier concentration locally results in refraction of the electron trajectories at the barrier interface. The refraction is symmetric with respect to the normal, which is identical to light refraction in optics. Thus far, asymmetric shift of the transverse momentum in electronic systems can only be induced by magnetic barriers that cause transverse Lorentz displacement or pseudo-magnetic barriers induced by strain in two-dimensional materials with particular lattice symmetry.

Weyl semimetals[6] constitute a recently discovered class of three-dimensional topological materials with linear band touching points at the Fermi level. A notable feature of 3D Weyl semimetals is that the dispersion around each Weyl point is generally tilted. It follows from that, due to symmetry constraints, Weyl points typically cannot reside on high-symmetry *k*-points. The low symmetry at the Weyl point location would usually allow a tilted spectrum, which is the case for almost all the Weyl semimetal materials identified to date. This is in contrast to the 2D graphene, in which the Dirac points are pinned at the high-symmetry *K* and *K'* points, forbidding possible tilt of spectrum. Then a natural question is how will the tilt affect the transport properties of Weyl fermions? Here, we show that asymmetric refraction and reflection of electrons can be achieved in Weyl semimetals with tilted energy dispersion by modulation of only electrical potential barrier. In other words, the perfect transmission angles exhibit a shift along the tilt direction of the energy dispersion, which is analogous to the transverse Lorentz displacement induced by magnetic barrier.

(*) email: can--yesilyurt@hotmail.com; elembaj@nus.edu.sg

A tilted Weyl fermion can be described by the low energy Weyl Hamiltonian with asymmetric velocities in three-dimension, i.e.,

$$H = V_0 + \sum_i \hbar k_i \left(\sigma^i v_i + w_i\right), \quad (1)$$

where $\sigma^i$'s are Pauli matrixes, $v_i$'s are velocities in three dimensions and $V_0$ is electrical potential. $w_i$'s describe the tilt of the energy dispersion in three dimensions. The tilt of the energy dispersion is a material dependent property. For instance, ~0.3 $v_F$ and ~0.1 $v_F$ tilted dispersion can be seen from the first principle calculations in TaP[7] and NbAs[8]. Therefore, we demonstrate the predicted conclusions with $w_y = 0.1 v_F$, a tilt strength which is within the observed range in various materials hosting Weyl fermions. By using Eq. (1), and considering $\gamma$ as the angle between $k_F$ and the $xy$ plane, $\phi$ as the azimuthal angle with respect to the $x$-axis, and assuming symmetric velocities that equal to the Fermi velocity $(v_i = v_F)$, the Fermi wave vector is $k_F = (E_F - V_0)/(\pm \hbar v_F + \hbar w_x \cos\gamma\cos\phi)$.

The required potential barrier configuration proposed in this work can be achieved by voltage gated doping or impurity doping, which are possible in current nanotechnology. For instance, in situ alkaline metal doping has been realized in three-dimensional Dirac semimetal $Cd_3As_2$[9]. Gate doping is also possible but it is somewhat constrained to thin film materials as the screening effect effectual in short distances. Electrostatic doping has been shown by solid electrolyte gating in ~50 nm-thick Dirac semimetal $Cd_3As_2$[10]. Besides, gate tunable carrier concentration has been also realized in ~14 nm-thick Weyl semimetal $WTe_2$[11]. These thicknesses are sufficient for the formation of bulk states in these Weyl and Dirac semimetals.

In this paper, we consider an infinite Weyl system where the contribution of the Fermi arc surface states to the conduction may be deemed negligible since the majority of contribution comes from the bulk states as shown in a thin film system of the Dirac semimetal $Na_3Bi$[12].

To investigate the electron tunneling transmission of a Weyl semimetal tilted along the $y$-direction, we calculate the angular dependence of tunneling probability in the case of Klein tunneling across an n-p-n junction with potential profile $V_{(x)} = V_0[\Theta(x) - \Theta(x - L)]$ as shown in Fig. 1[13]. Using the Hamiltonian (Eq. 1), the components of the wave functions of the system are found as

$$\psi_\pm \equiv \frac{1}{\sqrt{2}} e^{i\vec{k}\vec{r}} \begin{pmatrix} 1 \\ e^{i\phi} \sec\gamma(\pm 1 + \sin\gamma) \end{pmatrix} \equiv \begin{pmatrix} \psi_a \\ \psi_b^\pm \end{pmatrix} \quad (2)$$

By using matching of both components of the wave functions for incident, propagated and transmitted waves at the barrier interfaces, one can calculate the transmission probability across the system. The wave vectors outside of the barrier is expressed as $k_x = k_F \cos\gamma\cos\phi$, $k_y = k_F \cos\gamma\sin\phi$, $k_z = k_F \sin\gamma$ so that the $x$-component of wave vector within the barrier is $q_{x(w_y)} = \sqrt{(E_F - V_0 - \hbar k_y w_y)^2 - \hbar^2 v_F^2 (k_y^2 + k_z^2)}/\hbar v_F$. The angles of electrons propagation within the barrier $\theta = \tan^{-1}\left(\frac{k_y}{q_x}\right)$ and $\alpha = \tan^{-1}\left(\frac{k_z}{q_x}\cos\theta\right)$ can be calculated by considering the conservation of the transverse wave vectors $k_y$ and $k_z$ at the barrier interfaces.

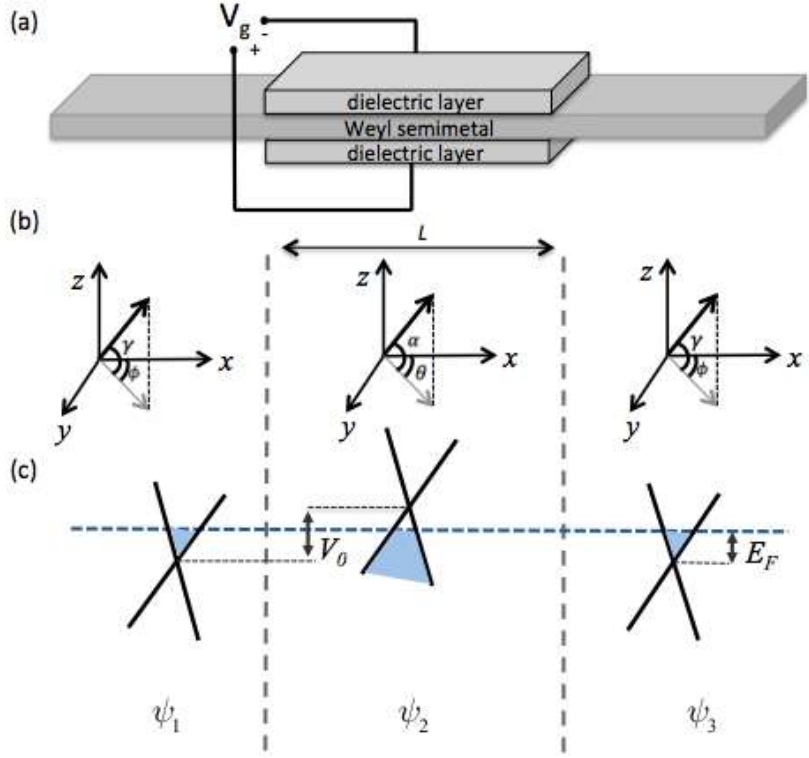

**Figure 1:** The schematic representation of a Weyl semimetal under the influence of an electrical barrier of width $L$, and height $V_0$. An n-p-n junction can be obtained in the case of $V_0 > E_F$, where $E_F$ is the Fermi level and shown by the dotted blue line.

In the non-tilted Weyl semimetal, the existence of potential barrier results in a symmetric transmission profile with respect to both $\phi = 0$ and $\gamma = 0$, as shown in Fig. 2 (a). This symmetry is broken when the energy dispersion is tilted along one of the transverse directions (y-direction in this example), as shown in Fig. 2 (b-d). The strength of the shift increases with increasing potential barrier height. The result is unusual since the application of electrical barrier does not give rise to an asymmetric tunneling profile with respect to normal incidence. To understand the origin of the anomalous shift, one can focus on the example shown in Fig. 3 (a), where the energy dispersion is tilted along the y-direction. The dashed and solid-lined circles represent the Fermi surfaces of Weyl fermions in the absence and presence of the electrical potential $V_0$ respectively. Conservation of energy and momentum at the barrier interface requires matching of the wave functions across the barrier, which is a prerequisite for non-zero tunneling probability. Simply, one may predict the range of wave vectors that allows electron transmission by considering the conservation and hence, overlap of transverse wave-vectors. In the case of $\phi > \phi_{a,b}$, $q_x$ becomes imaginary and total reflection occurs. The shaded angles in Fig. 3 show the allowed range of incident vectors for transmission at the first barrier interface.

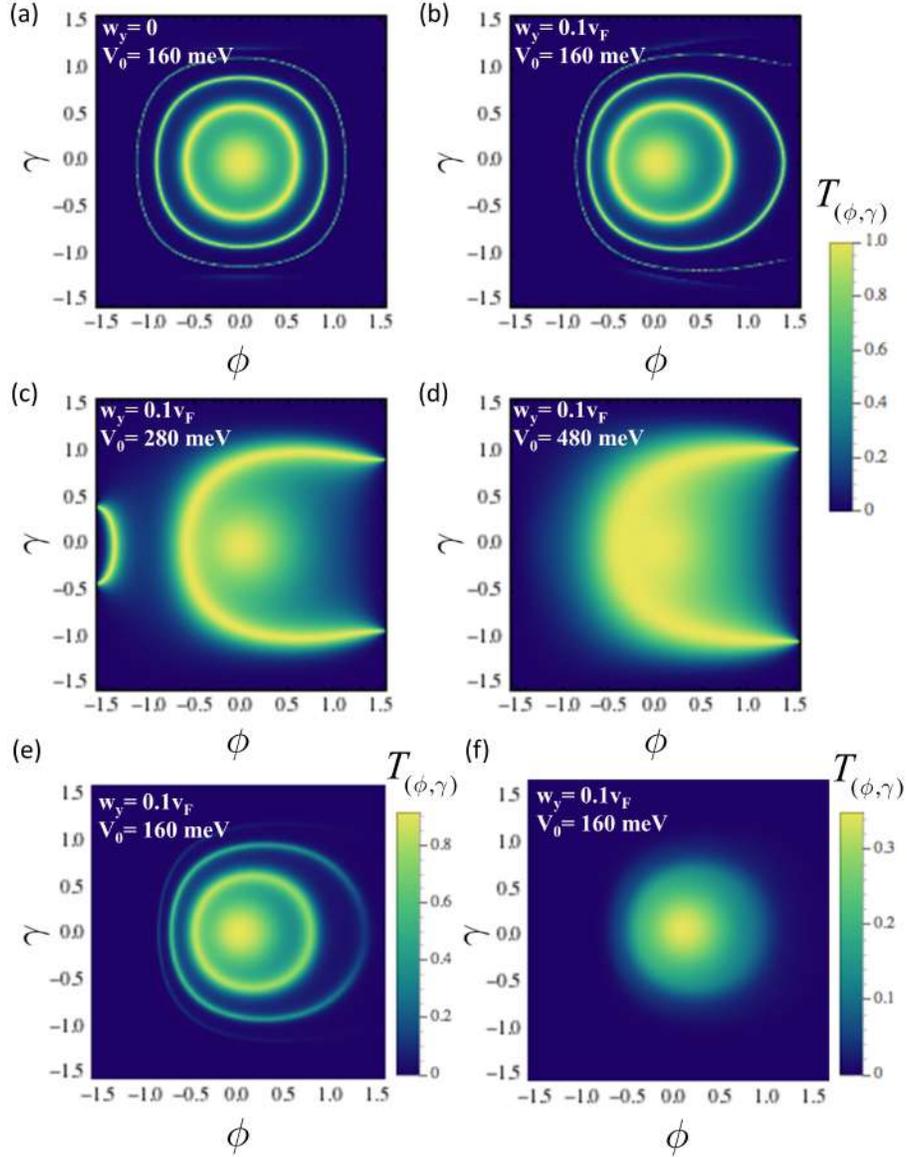

**Figure 2:** The angular dependence of tunneling probability for n-p-n junction of a Weyl semimetal with (a) non-tilted and (b-d) tilted energy dispersion by $w_y = 0.1 \, v_y$ under the influence of applied potential $V_0$ on the central region. (e) and (f) show the same system configurations with (b) in the case of disorder scattering. (e) shows the effect of weak disorder (ξ = 1 meV), while (f) shows strong disorder effect (ξ = 10 meV). Fermi energy $E_F = 82.6$ meV, the barrier length $L = 100$ nm for all configurations, and the incident angles are in radians.

In Fig. 3 (a), although there is no effect that breaks the symmetry of momentum such as magnetic vector potential or strain gauge[4], the angle range between $\phi_a$ and $\pi/2$ leads to imaginary $q_x$, thus the electrons whose incident angle falls within this range would experience total reflection at the barrier interface.

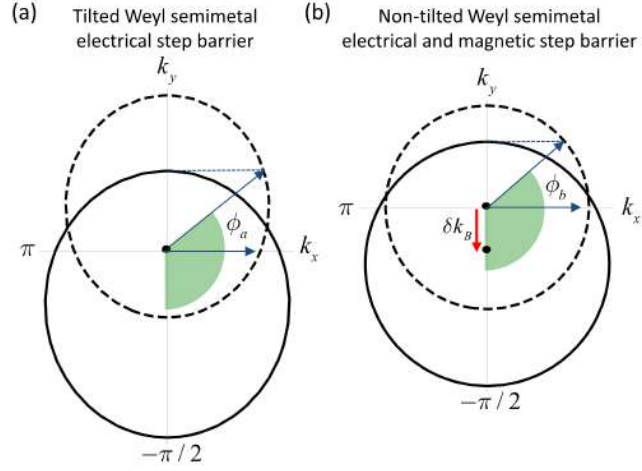

**Figure 3:** The Fermi surfaces of Weyl fermions belong to different regions shown in Fig. 1. Dashed (straight) circle represents the Fermi surface for Weyl fermion in the absence (presence) of electric potential $V_0 = $ 176 meV, where Fermi energy $E_F = 82.6$ meV. (a) illustrates a Weyl system where $w_y = -0.4\, v_F$. (b) shows a Weyl system where $w_y = 0$, $\delta k_B \neq 0$. For clarity of the Fermi surfaces, we set $k_z = 0$.

It can be analytically found that, in the case of $w_y \neq 0$, and $k_z = 0$ (Fig. 3 (a)), the upper limit of the incident angle with non-zero transmission is

$$\phi_a = \sin^{-1}\left(-\frac{v_F(V_0 - E_F)}{E_F v_F - V_0 w_y}\right) \qquad (3)$$

The case $w_y \to 0$ implies that $\phi_a \to \pi/2$ when $V_0 = 2E_F$, which means the anomalous shift vanishes in a non-tilted band structure. Note that Eqs. (3) and (4) are valid in the range $E_F < V_0 < 2E_F v_F/(v_F + w_y)$ and express the limit for positive $\phi$, while the limit for the negative $\phi$ is not affected by the transverse shift and always equal to $-\pi/2$ for the configurations shown in Fig. 3.

The shifted transmission profiles in Fig. 2 are calculated by considering a single node located at **K** described by $H = v_F \vec{k}.\vec{\sigma} + \vec{w}.\vec{k}$. In the simplest Weyl semimetal case, another Weyl node must appear at –**K** with opposite chirality due to the inversion symmetry, which is described by $H = -v_F \vec{k}.\vec{\sigma} - \vec{w}.\vec{k}$. At –**K**, we found that the transmission profile shows identical pattern but reversed with respect to $\phi = 0$. In such a system, the anomalous shift would cause deflections in the opposite directions at the barrier interface for the two valleys according to their respective chirality. This effect is analogous to the valley dependent deflections at the interface of strained and unstrained regions in two-dimensional hexagonal lattices such as graphene, which is commonly used in valleytronics to lift the valley degeneracy [14,15].

In order to evaluate the robustness of the proposed effect against disorders, we consider the standard real space finite-difference NEGF formalism[16], and represent scattering potentials that destroy phase coherence by means of Buttiker probe (virtual lead) attached to every lattice sites in the central region. The amount of decoherence is controlled by the coupling strength of each lattice site to its virtual lead, which is characterized by the energy broadening (i.e., ξ) or inverse of the scattering lifetime of the site state. As an example, we have re-calculated the angular dependence of the transmission probability in a system with the same configurations as that shown in Fig. 2 (b) for a system without scattering. As shown in Fig. 2 (e) and (f), disorder scattering causes a reduction in the total transmission and conductance. Besides, the perfect transmission angles lose their sharpness due to the dephasing effect. However, we note that crucially the asymmetric profile of the transverse momentum shift still holds true, even though the conductance is reduced by ∼10% in the case of ξ = 1 meV. From the result shown in the Fig. 2 (f), one can predict that only in the case of very strong disorder (ξ = 10 meV) which can reduce the conductance by ∼70%, would the effects of the anomalous shift be destroyed.

Previously, such an asymmetric momentum shift at the barrier interface was analyzed only in the presence of step magnetic or strain gauge potential[4,14,17]. One can thus compare this system shown in Fig. 3 (a) with non-tilted Weyl system in the presence of a step magnetic vector potential where the Fermi surfaces of two different regions are shown in Fig. 3 (b). The shift of the Fermi surface due to the magnetic vector potential, $\delta k_B$ can generate a mismatch of $k_y' = k_y + \delta k_B$ which may lead to constraints on the allowable wave-vectors for transmission. In this case of non-tilted Weyl system and applied magnetic step vector potential $A_y = B_0 l_B \Theta(x) \hat{y}$, where $l_B = \sqrt{\frac{h}{|e|B_0}}$, the upper limit of the incident angle that can lead transmission is found as

$$\phi_b = \sin^{-1}\left(-\frac{V_0 - E_F + \eta v_F \sqrt{|B_0|he}}{E_F}\right), \tag{4}$$

where $\eta = \text{sign}(B_z)$. Owing to the fact that the angles $\phi_a$ and $\phi_b$ is a measure of the transverse shift arising from the dispersion tilt and $B$-field, respectively, by equating them we can find the effective $B$-field corresponding to a particular tilt strength. Numerically, it can be found that $\phi_a \approx \phi_b \approx 0.7$ for the configurations shown in Fig. 3 (a) and (b), where $\delta k_B = eA_y/h \approx 6.7 \times 10^7$ m$^{-1}$. That means $B_0 = \frac{h \delta k_B}{e} \approx 3$ T delta-function magnetic field is required to obtain the same shift of critical angle caused by $w_y = 0.1\, v_F$ and $V_0 = 176$ meV.

Likewise, in the case of n-p-n junction shown in Fig. 1, the equivalent transverse momentum shift generated by coupling of the tilt strength and electrical potential can be approximately derived by considering the average of shift at the maximum and minimum points of $k_y$. The magnitude of the delta function pseudo-magnetic field generated by the application of an electrical potential is given by

$$B_0 \approx \frac{V_0^2 w_y^2}{eh\left(w_y^2 - v_F^2\right)^2}. \tag{5}$$

The results shown in Fig. 4 for the delta B-field barrier with strength given in Eq. (5) are almost identical with that of the n-p-n junction of tilted Weyl semimetal without magnetic barrier (see Fig. 2).

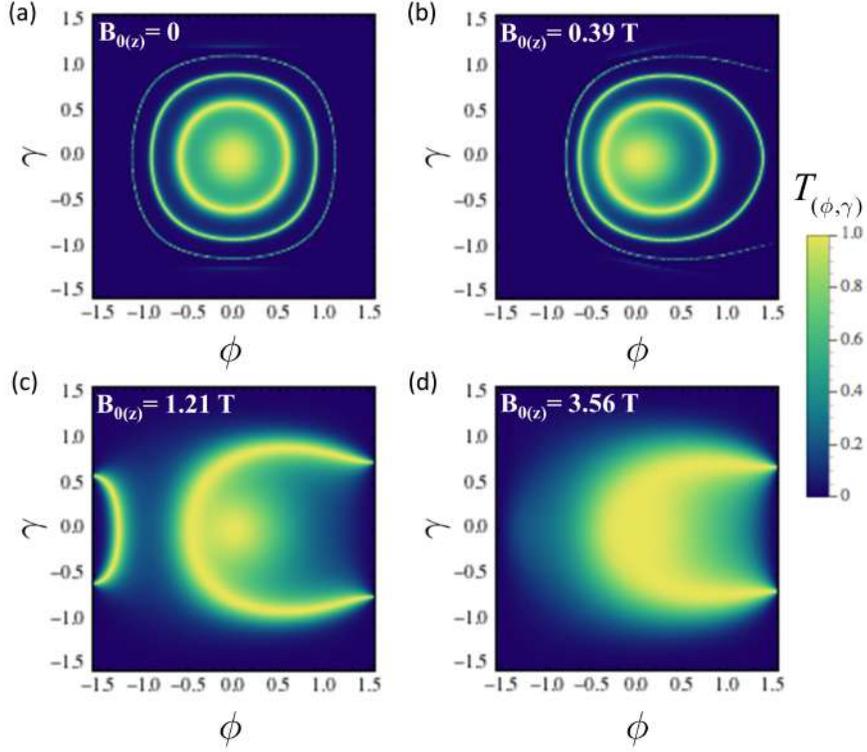

**Figure 4:** The angular dependence of tunneling probability for n-p-n junction of a Weyl semimetal with conventional non-tilted energy dispersion under the influence of different magnetic field strengths $B_z$ and applied potential $V_0$ on the central region whose strengths are the same with the configurations in Fig. 2. Fermi energy $E_F = 82.6$ meV, the barrier length $L = 100$ nm for all configurations, and the incident angles are in radians.

Although the delta-function magnetic field model is useful to obtain the equivalent pseudo-magnetic barrier strength (Eq. 5), the magnetic field profile is unlikely to be very sharp in reality. Therefore, we consider the more realistic magnetic field profile proposed in Ref. [18], where the gauge potential can be assumed to possess smoothly varying profile in the range $d_x$, as illustrated in Fig. 5 (a). For comparison, the height of the magnetic gauge barrier has been kept the same as that in the delta-function model. We applied the standard real space finite-difference NEGF formalism[16] to calculate the angular dependence of transmission probability for the system. As can be seen in Fig. 5 (b), the transmission profile is hardly affected by the finite spatial width $d_x$ of the magnetic field region. We can conclude that the transmission profile, especially the allowed transmission angles as discussed in the manuscript, does not depend on the spatial extent of the magnetic barrier, as much as the magnetic gauge barrier height, which is the factor we focused on in this work.

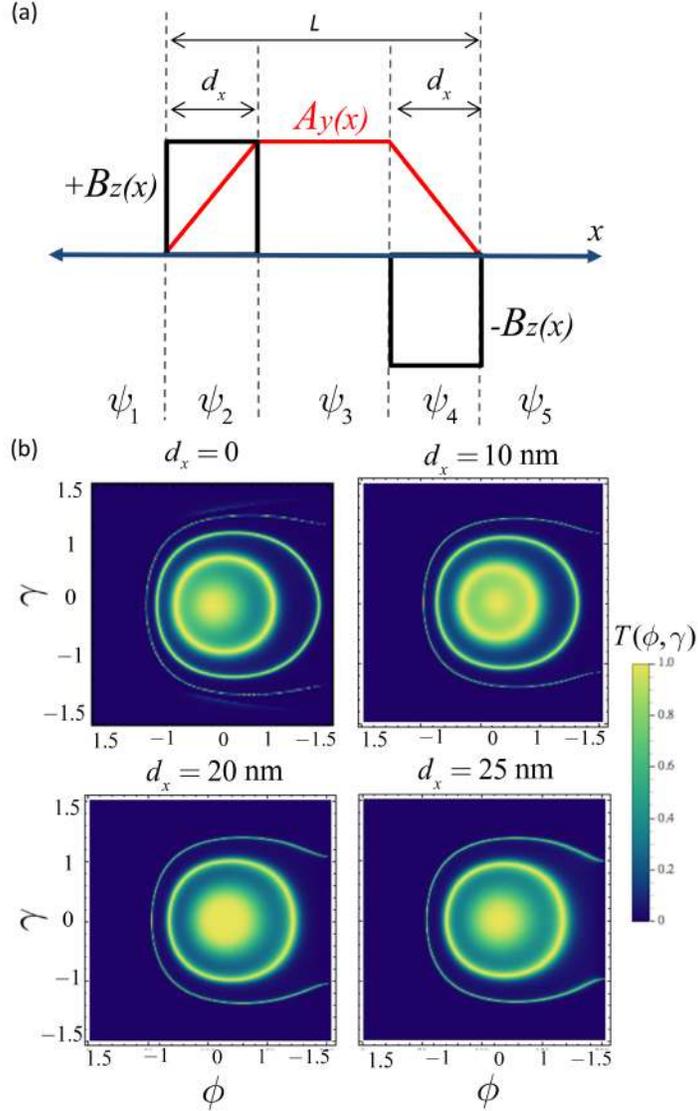

**Figure 5**. (a) The schematic illustration of magnetic barrier described in the text. (b) The angular dependence of the transmission probability, where all the system parameters, i.e., not specified above, are the same with Fig. 4 (b). For comparison, the result with the delta-function approximation ($d_x = 0$) is also plotted.

Our findings clearly show that an n-p-n junction of Weyl semimetal with tilted energy dispersion may exhibits a transmission profile that is almost identical to that of a conventional (non-tilted) Weyl in the presence of a magnetic barrier. The similarity of these two different effects can be intuitively understood by analyzing the band structure in each case. As shown in the above model, the magnetic gauge potential shifts the Weyl cones in *k*-space. Also, the tilt of the Weyl fermion shifts the Weyl cones, but this arises in a momentum and energy-dependent way. In the presence of a potential energy difference between two regions of Weyl semimetal, the tilt causes an asymmetric mismatch of Weyl cones at a particular energy, while for non-tilted systems, the mismatch would be symmetric. Although these two effects may give rise to a similar outcome regarding the electron transport, the tilt-dependent pseudo-gauge field crucially differs from magnetic barriers since the magnetic fields would shift the whole Weyl cone by the same amount while the shift due to the coupling of the tilt and potential energy gradient would leave the Weyl point unchanged in *k*-space. Thus, when the wave vector shift induced by magnetic gauge potential $eA_y/\hbar$ is larger than the transverse wave vector $k_y$, the barrier would forbid transmission of normally incident electrons at zero Kelvin, an outcome which is impossible to obtain with only electrical barriers in tilted systems. Therefore, for the case of $B_z > (E_F - V_0)^2/(e \hbar v_F^2)$, there would be a clear difference in the transmission for a tilted system under an electrical potential barrier, and that of a non-tilted system under a magnetic barrier.

As a conclusion, the application of electrical potential in Dirac and Weyl semimetals change the carrier concentration but does not break the symmetry of the momentum. Here, we showed that the application of electrical potential is able to shift momentum, in addition to changing the carrier concentration. Our analysis reveals signature features in the transport characteristics that may help identify materials having the tilted Weyl dispersion. The findings may be utilized to engineer electro-optic applications such as electron beam collimation, electron lenses, and suppression of tunneling transmission controlled by electrical gate bias.

## ACKNOWLEDGEMENT

The authors would like to acknowledge the MOE Tier II grant MOE2013-T2-2-125 (NUS Grant No. R-263-000-B10-112), and the National Research Foundation of Singapore under the CRP Program "Next Generation Spin Torque Memories: From Fundamental Physics to Applications" NRF-CRP9-2013-01 for financial support.